\documentclass[12pt]{article}
\usepackage{amsfonts}
\usepackage{amssymb}

\def\hybrid{\topmargin -20pt    \oddsidemargin 0pt
        \headheight 0pt \headsep 0pt
        \textwidth 6.35in       
        \textheight 9.25in       
        \marginparwidth .875in
        \parskip 5pt plus 1pt   \jot = 1.5ex}

\hybrid

\def\baselinestretch{1.2}

\catcode`\@=11

\def\marginnote#1{}
\newcount\hour
\newcount\minute
\newtoks\amorpm
\hour=\time\divide\hour by60
\minute=\time{\multiply\hour by60 \global\advance\minute by-\hour}
\edef\standardtime{{\ifnum\hour<12 \global\amorpm={am}%
        \else\global\amorpm={pm}\advance\hour by-12 \fi
        \ifnum\hour=0 \hour=12 \fi
        \number\hour:\ifnum\minute<10 0\fi\number\minute\the\amorpm}}
\edef\militarytime{\number\hour:\ifnum\minute<10 0\fi\number\minute}

\def\draftlabel#1{{\@bsphack\if@filesw {\let\thepage\relax
   \xdef\@gtempa{\write\@auxout{\string
      \newlabel{#1}{{\@currentlabel}{\thepage}}}}}\@gtempa
   \if@nobreak \ifvmode\nobreak\fi\fi\fi\@esphack}
        \gdef\@eqnlabel{#1}}
\def\@eqnlabel{}
\def\@vacuum{}
\def\draftmarginnote#1{\marginpar{\raggedright\scriptsize\tt#1}}

\def\draft{\oddsidemargin -.5truein
        \def\@oddfoot{\sl preliminary draft \hfil
        \rm\thepage\hfil\sl\today\quad\militarytime}
        \let\@evenfoot\@oddfoot \overfullrule 3pt
        \let\label=\draftlabel
        \let\marginnote=\draftmarginnote
   \def\@eqnnum{(\theequation)\rlap{\kern\marginparsep\tt\@eqnlabel}%
\global\let\@eqnlabel\@vacuum}  }

\def\preprint{\twocolumn\sloppy\flushbottom\parindent 2em
        \leftmargini 2em\leftmarginv .5em\leftmarginvi .5em
        \oddsidemargin -.5in    \evensidemargin -.5in
        \columnsep .4in \footheight 0pt
        \textwidth 10.in        \topmargin  -.4in
        \headheight 12pt \topskip .4in
        \textheight 6.9in \footskip 0pt
        \def\@oddhead{\thepage\hfil\addtocounter{page}{1}\thepage}
        \let\@evenhead\@oddhead \def\@oddfoot{} \def\@evenfoot{} }

\def\numberbysection{\@addtoreset{equation}{section}
        \def\theequation{\thesection.\arabic{equation}}}

\def\underline#1{\relax\ifmmode\@@underline#1\else
        $\@@underline{\hbox{#1}}$\relax\fi}

\def\titlepage{\@restonecolfalse\if@twocolumn\@restonecoltrue\onecolumn
     \else \newpage \fi \thispagestyle{empty}\c@page\z@
        \def\thefootnote{\fnsymbol{footnote}} }

\def\endtitlepage{\if@restonecol\twocolumn \else \newpage \fi
        \def\thefootnote{\arabic{footnote}}
        \setcounter{footnote}{0}}  

\catcode`@=12
\relax

\def\figcap{\section*{Figure Captions\markboth
        {FIGURECAPTIONS}{FIGURECAPTIONS}}\list
        {Figure \arabic{enumi}:\hfill}{\settowidth\labelwidth{Figure
999:}
        \leftmargin\labelwidth
        \advance\leftmargin\labelsep\usecounter{enumi}}}
 \relax
\def\tablecap{\section*{Table Captions\markboth
        {TABLECAPTIONS}{TABLECAPTIONS}}\list
        {Table \arabic{enumi}:\hfill}{\settowidth\labelwidth{Table
999:}
        \leftmargin\labelwidth
        \advance\leftmargin\labelsep\usecounter{enumi}}}
 \relax
\def\reflist{\section*{References\markboth
        {REFLIST}{REFLIST}}\list
        {[\arabic{enumi}]\hfill}{\settowidth\labelwidth{[999]}
        \leftmargin\labelwidth
        \advance\leftmargin\labelsep\usecounter{enumi}}}
 \relax
\makeatletter
\newcounter{pubctr}
\def\publist{\@ifnextchar[{\@publist}{\@@publist}}
\def\@publist[#1]{\list
        {[\arabic{pubctr}]\hfill}{\settowidth\labelwidth{[999]}
        \leftmargin\labelwidth
        \advance\leftmargin\labelsep
        \@nmbrlisttrue\def\@listctr{pubctr}
        \setcounter{pubctr}{#1}\addtocounter{pubctr}{-1}}}
\def\@@publist{\list
        {[\arabic{pubctr}]\hfill}{\settowidth\labelwidth{[999]}
        \leftmargin\labelwidth
        \advance\leftmargin\labelsep
        \@nmbrlisttrue\def\@listctr{pubctr}}}
 \relax
\makeatother
\newskip\humongous \humongous=0pt plus 1000pt minus 1000pt

\newif\ifdtup

\relax

\def\be{\begin{equation}}
\def\ee{\end{equation}}
\def\ba{\begin{eqnarray}}
\def\ea{\end{eqnarray}}

\def\del{\partial}

\def\r{\rho}
\def\a{\alpha}

\def\b{\beta}

\def\g{\gamma}
\def\G{\Gamma}
\def\d{\delta}

\def\e{\epsilon}

\def\m{\mu}

\def\om{\omega}

\def\l{\lambda}
\def\L{\Lambda}
\def\s{\sigma} 
\def\S{\Sigma}

\def\cN{{\cal N}}

\def\bs{\bigskip}

\def\qq{\qquad}

\def\IR{\relax{\rm I\kern-.18em R}}

\def \ha {{1\over 2}}

\def \ov {\over}

\def\IR{\relax{\rm I\kern-.18em R}}
\def\inv{^{\raise.15ex\hbox{${\scriptscriptstyle -}$}\kern-.05em 1}}

\def\hi{{\hat i}}
\def\hj{{\hat j}}
\def\hk{{\hat k}}

\begin{document}

\newcommand{\beq}{\begin{equation}}
\newcommand{\eeq}[1]{\label{#1}\end{equation}}
\newcommand{\ber}{\begin{eqnarray}}
\newcommand{\eer}[1]{\label{#1}\end{eqnarray}}
\newcommand{\eqn}[1]{(\ref{#1})}
\begin{titlepage}
\begin{center}

\hfill NEIP-02-002\\
\vskip -.1 cm
\hfill hep--th/0202135\\

\vskip .5in

{\LARGE An eight--dimensional approach to $G_2$ manifolds}

\vskip 0.4in

{\bf Rafael Hern\'andez$^1$}\phantom{x} and\phantom{x}
 {\bf Konstadinos Sfetsos}$^2$ 
\vskip 0.1in

${}^1\!$
Institut de Physique, Universit\'e de Neuch\^atel\\
Breguet 1, CH-2000 Neuch\^atel, Switzerland\\
{\footnotesize{\tt rafael.hernandez@unine.ch}}

\vskip .2in

${}^2\!$
Department of Engineering Sciences, University of Patras\\
26110 Patras, Greece\\
{\footnotesize{\tt sfetsos@mail.cern.ch, des.upatras.gr}}\\

\end{center}

\vskip .3in

\centerline{\bf Abstract}

We develop a systematic approach to $G_2$ holonomy manifolds with an $SU(2)
\times SU(2)$ isometry using maximal eight-dimensional gauged supergravity 
to describe D6-branes wrapped on deformed three-spheres. 
A quite general metric ansatz that generalizes the celebrated 
Bryant--Salamon metric involves nine functions. We show that only six
of them are the independent ones and derive 
the general first order system of differential equations that they obey.
As a byproduct of our analysis, we generalize the notion of the twist that
relates the spin and gauge connections in a way that 
involves non-trivially the scalar fields.

\noindent

\vskip .4in
\noindent

\end{titlepage}
\vfill
\eject

\def\baselinestretch{1.2}


\baselineskip 20pt

Compactifications of M-theory on manifolds of exceptional holonomy 
have recently attracted 
great attention, mostly as a consequence of their relation 
to minimally supersymmetric gauge 
theories. Four dimensional ${\cal N}=1$ supersymmetry (in Minkowski space)
requires the internal seven manifold to have 
$G_2$ holonomy. 
But $G_2$ holonomy also appears in the geometric dual description of the 
large $N$ limit of four dimensional gauge theories with four supercharges: the conjectured 
duality between D6-branes on the deformed conifold and a type IIA geometry with RR flux on the 
resolved conifold in \cite{Vafa} was better understood in terms of M-theory on a seven 
manifold of $G_2$ holonomy \cite{Acharya}, 
where it corresponds to a flop transition \cite{AMV}. Extensions of 
this duality and construction of new metrics from diverse approaches have 
revived the study of compactifications on manifolds of exceptional 
holonomy \cite{Cachazo}-\cite{AOG}. 
  
The number of known complete metrics of $G_2$ holonomy is still quite reduced. It is therefore of 
great interest to obtain new metrics of $G_2$ holonomy in order to improve our understanding of the above dualities 
and compactifications. The aim of this letter is to elaborate on a gauged supergravity approach to the 
systematic construction of manifolds of $G_2$ holonomy. 
  
Branes wrapped on supersymmetric cycles have also been lately quite extensively studied within the framework of gauged 
supergravity as a promising candidate to gravity 
duals of field theories with low supersymmetry \cite{MN1}-\cite{DiV}. In \cite{EN} a configuration 
of D6-branes wrapping special Lagrangian 3-spheres was considered as a gravity dual of four dimensional 
field theories with ${\cal N}=1$ supersymmetry. The lift to eleven dimensions of the eight 
dimensional solution describing the deformation on the worldvolume of the wrapped branes was there shown to correspond 
to one of the known metrics of $G_2$ holonomy \cite{BS}. In this letter we 
will show how gauged supergravity in 
eight dimensions provides a natural framework to construct general metrics of $G_2$ holonomy by allowing 
deformations on the 3-cycle. We will derive the conditions to guarantee $G_2$ holonomy on a seven manifold metric of the form 
\[
ds_7^2 = dr^2 + \sum_{i=1}^3 a_i^2 \s_i^2 +
\sum_{i=1}^3 b_i^2(\S_i + c_i \s_i)^2\ ,
\]
where, as it will become clear from our analysis below, 
only six of the nine functions involved in this general 
metric are independent. 

In what follows we will briefly review some relevant facts about eight dimensional supergravity. We will then 
construct the equations describing a supersymmetric configuration corresponding to a set of D6-branes 
wrapped on a deformed 3-sphere. The lift to eleven dimensions of this configuration 
will prove to be a seven manifold of $G_2$ holonomy with $SU(2) \times SU(2)$ isometry which includes 
some of the proposed ansatzs in the literature.
  
Maximal gauged supergravity in eight dimensions was constructed by Salam and Sezgin \cite{Salam}
through Scherk--Schwarz compactification of eleven dimensional supergravity on 
an $SU(2)$ group manifold. The field content in the gravity sector of 
the theory consists of the metric $g_{\mu \nu}$, a dilaton 
$\Phi$, five scalars given by a unimodular $3 \times 3$ matrix $L_{\alpha}^{i}$ in the coset 
$SL(3, {\bf R})/SO(3)$ and an $SU(2)$ 
gauge potential $A_{\mu}$.\footnote{The fields arising from reduction of the eleven dimensional three-form are a scalar, 
three vector fields, three two-forms and a three-form. However, 
we will only consider pure gravitational solutions of the eleven dimensional theory, so that 
all these fields can be set to zero.} 
In addition, on the fermion side we have 
the pseudo--Majorana spinors $\psi_{\mu}$ and $\chi_i$.
  
The Lagrangian density 
for the bosonic fields is given, in $\kappa=1$ units, by 
\begin{equation} 
{\cal L} = \frac {1}{4} R - \frac {1}{4} e^{2 \Phi} F_{\mu \nu}^{\a} F^{\mu \nu \; \b} g_{\a \b}- 
\frac {1}{4} P_{\mu \; ij} P^{\mu \; ij} - \frac {1}{2} (\partial_{\mu} \Phi)^2 - 
\frac {g^2}{16} e^{-2 \Phi} ( T_{ij} T^{ij} - \frac {1}{2} T^2) \, ,
\end{equation}
where $F_{\mu \nu}^{\a}$ is the Yang--Mills field strength.
  
Supersymmetry is preserved by bosonic solutions to the equations 
of motion of eight dimensional 
supergravity if the supersymmetry variations for the 
gaugino and the gravitino vanish. These are, respectively, given by 
\be
\delta \chi_i  = 
 \frac {1}{2} (P_{\mu \; ij} + \frac {2}{3} \delta_{ij} \partial_{\mu} 
\Phi) \hat{\Gamma}^{j} 
\Gamma^{\mu} \epsilon - \frac {1}{4} e^{\Phi} F_{\mu \nu \; i} 
\Gamma^{\mu \nu} \epsilon 
- \frac {g}{8} e^{-\Phi} (T_{ij} - \frac {1}{2} \delta_{ij} T) 
\epsilon^{jkl} \hat{\Gamma}_{kl} \epsilon = 0 \, 
\label{susy}
\ee
and 
\be
\delta \psi_{\gamma}  = 
{\cal D}_{\gamma} \epsilon + \frac {1}{24} e^{\Phi} F_{\mu \nu}^{i} 
\hat{\Gamma}_i ( \Gamma_{\gamma}^{\: \: \mu \nu} 
- 10 \delta_{\gamma}^{\, \mu} \Gamma^{\nu}) \epsilon 
- \frac {g}{288} e^{- \Phi} \epsilon_{ijk} 
\hat{\Gamma}^{ijk} \Gamma_{\gamma} T \epsilon = 0 \ .
\label{susyg}
\ee
The covariant derivative is
\begin{equation}
{\cal D}_{\mu} \epsilon = \partial_{\mu} \epsilon + \frac {1}{4} \omega_{\mu}^{ab} \Gamma_{ab} 
\epsilon + \frac {1}{4} Q_{\mu \; ij} \hat{\Gamma}^{ij} \epsilon \, ,
\end{equation}
where $P_{\mu \, ij}$ and $Q_{\mu \, ij}$ are, respectively, 
the symmetric and antisymmetric quantities entering 
the Cartan decomposition of the $SL(3,{\bf R})/SO(3)$ coset, defined through
\begin{equation}
P_{\mu \, ij} + Q_{\mu \, ij} \equiv L_i^{\alpha} 
( \partial_{\mu} \delta_{\alpha}^{\, \beta} 
- g \, \epsilon_{\alpha \beta \gamma} A^{\gamma}_{\mu}) L_{\beta \; j} \, ,
\label{pq}
\end{equation}
and $T_{ij}$ is the $T$-tensor defining the potential energy associated to the scalar fields,
\begin{equation}
T^{ij} \equiv L^{i}_{\alpha} L^{j}_{\beta} \delta^{\alpha \beta} \, ,
\end{equation}
with $T \equiv T_{ij} \delta^{ij}$, and 
\be 
L_{\a}^{i} L^{\a}_j = \delta^{i}_j\ , \: \: \: \: L_{\a}^{i} L_{\b}^{j} 
\d_{ij} = g_{\a \b}\ , 
\: \: \: \: L^{i}_{\a} L^{j}_{\b} g^{\a \b} = \d^{ij}\ .
\ee
As usual, curved directions are labeled by greek indices, while 
flat ones are labeled by latin, and $\mu, a = 0,1, \ldots, 7$ are spacetime coordinates, 
while $\alpha, i = 8,9,10$ are in the group manifold. Note also that upper indices in the gauge field, 
$A_{\mu}^{\alpha}$, are always curved.   
  
We will turn on scalars in the diagonal
\be
L^i_{\a} = \hbox {diag} ( e^{\l_1}, e^{\l_2}, e^{\l_3})\ ,
\qq {\l_1}+{\l_2}+{\l_3}=0 \ ,
\label{lia}
\ee
and in order to describe the worldvolume of the wrapped D6-branes 
on the deformed 3-cycle we will choose a metric ansatz of the 
form\footnote{We should 
note that deformation of the 3-cycle requires the existence of 
non-trivial scalars on the coset manifold.}
\be
ds_8^2 = \a_1^2 \s_1^2 + \a_2^2 \s_2^2 + \a_3^2 \s_3^2 + e^{2f} 
ds_{1,3}^2 + d\r^2 \, .
\label{s8}
\ee
All four functions $\a_i$, $f$ as well as the scalars $\l_i$ and the dilaton
$\Phi$ depend only on $\rho$, and the left-invariant
Maurer--Cartan $SU(2)$ 1-forms satisfy
\be
d\s_i= \ha \e_{ijk} \s_j\wedge \s_k\ .
\ee
In this basis we also expand the gauge field 1-forms as $A^\a=A^\a_i \s^i$,
with components $A^\a_i$ that depend only on the variable $\r$.
For simplicity, the four-dimensional metric $ds^2_{1,3}$ 
will be taken to be the Minkowski 
metric, but our analysis can be easily extended to Ricci-flat metrics.
  
We will represent the $32\times 32$ gamma matrices in 11 dimensions as
\begin{equation}
\Gamma^{a} = \gamma^{a} \times {\bf 1}_2, \: \: \: \: \: \: \: \: \hat{\Gamma}^{i} = \gamma_9 \times \tau^{i} \, ,
\end{equation}
where the $\g_a$'s denote the $16\times 16$ gamma matrices in
8 dimensions 
and as usual $\gamma_9 = i \gamma^0 \gamma^1 \ldots \gamma^7$, 
so that $\gamma_9^2 = {\bf 1}$. Also $\tau^{i}$ 
are Pauli matrices. It will prove useful to introduce 
$\Gamma_9 \equiv \frac {1}{6i} \epsilon_{ijk} \hat{\Gamma}^{ijk} 
= -i \hat{\G}_1 \hat{\G}_2 \hat{\G}_3 = 
\gamma_9 \times {\bf 1_2}$.
  
Within this ansatz, the only consistent way to obtain non-trivial solutions to 
the Killing spinor equations is to impose on the spinor $\e$ the projections
\be
\G_{ij} \e = - \hat{\G}_{ij} \e\ , \: \: \: \: \G^7 \e = - i \G^9 \e\ .
\label{projections}
\ee
The first of these projections relates the two $SU(2)$ algebras
obeyed separately by the sets of generators 
$\{ \G_{ij} \}$ and $\{ \hat\G_{ij} \}$ and consequently
the ``spacetime'' and internal deformed 3-spheres.
It also states that only singlets of the diagonal $SU(2)_D$ of the 
tensor product of the two $SU(2)$'s are allowed.
We emphasize that simple algebraic considerations reveal that 
the only allowed coefficient in a relation of the form 
$\G_{ij} \e = \l \, \hat{\G}_{ij} \e$ is $\l=-1$. 
We also note that among the possible pairs 
$\{ij\}=\{12,23,31\}$ only two are independent. 
Therefore the projections (\ref{projections}) represent three conditions 
in total, thus reducing the number of 
supersymmetries to $32/2^3=4$.\footnote{All conditions 
in \eqn{projections}
can be cast in the form 
\ba
(\G_7 \hat \G_i + \ha \e_{ijk}\G_{jk})\e=0\ .
\nonumber
\ea
}
In the forthcoming derivation of the equations, the relations
\ba
&& \G_i \hat{\G}_j \e  =  \hat{\G}_i \G_j \e \ , \qquad i \neq j \ , 
\nonumber \\
&& \G_1 \hat{\G}_1 \e = \G_2 \hat{\G}_2 \e = \G_3 \hat{\G}_3 \e \ ,
\ea
which can be readily derived from \eqn{projections}, will also be 
useful.
  
When we wrap the D6-branes on the 3-cycle the 
$SO(1,6) \times SO(3)_R$ symmetry group of the unwrapped 
branes is broken to $SO(1,3) \times SO(3) \times SO(3)_R$. The worldvolume of the brane 
will support covariantly constant spinors after some twisting or mixing of the spin and gauge connections. 
In the presence of scalars this twisting can not be 
simply performed through a direct identification of the spin and gauge connections. 
As detailed in the appendix 
the gauge field is defined through the generalized 
twist\footnote{In the absence of scalar fields, 
with $L_{\a}^{i} = \delta_{\a}^{i}$, and with no deformation 
of the 3-sphere, 
the gauge field reduces simply to $A^{i}_i = - \frac {1}{2g}$ \cite{EN}.}  
\be
\frac {1}{g} \frac {\omega^{23}_1}{\a_1} 
+  \frac {A^1_1}{\a_1} \cosh \l_{23} + 
\frac {A^2_2}{\a_2} \sinh \l_{31} - \frac {A_3^3}{\a_3} \sinh \l_{12} = 0
\label{twist}
\ee
and cyclic in $1,2,3$, so that the solution is
\be
A_1^1 = \frac {\a_1}{g} \Big[ - \frac {\omega_1^{23}}{\a_1} \cosh \l_{23} 
+ e^{\l_{21}} \sinh \l_{31} 
\frac {\omega^{31}_2}{\alpha_2} - e^{\l_{31}} \sinh \l_{12} 
\frac {\omega^{12}_3}{\a_3} \Big] \, ,
\label{A1}
\ee 
and so on for $A^2_2$ and $A^3_3$. We have used the notation 
\be \omega^{jk}_i= \e_{ijk}\frac {\a_j^2 + \a_k^2 - \a_i^2}{2\a_j \a_k}\ ,
\label{sppin}
\ee
for the components of the spin connection along the 3-sphere expanded as 
$\om^{jk}=\om^{jk}_i \s_i$, 
and $\l_{ij}=\l_i-\l_j$. 
We see that, in general, the relation between the spin connection and the 
gauge field involves in a rather complicated way the scalar fields.
     
A detailed account of the computations required to derive the equations 
obeyed by the various fields is given in the appendix. Here we just collect 
the results. 
From the gaugino variation in \eqn{susy} one obtains the equations 
obeyed by the dilaton 
\be
{d\Phi\ov d\rho} = {1\ov 2} e^\Phi \left( \frac {e^{\lambda_1}}{\a_2 \a_3} F^{1}_{23} + \frac {e^{\lambda_2}}{\a_3\a_1} F^2_{31} 
+ \frac {e^{\lambda_3}}{\a_1 \a_2} F^{3}_{12} \right)
+ {g\ov 8} e^{-\Phi} 
(e^{2\l_1} + e^{2\l_2} + e^{2\l_3})\ .
\label{dilaton}
\ee
and by the scalars, 
\be
{d\l_1\ov d\rho} = \frac {e^{\Phi}}{3} \left( 2 \frac {e^{\lambda_1}}{\a_2 \a_3} F^{1}_{23} - 
\frac {e^{\lambda_2}}{\a_3\a_1} F^2_{31} - 
\frac {e^{\lambda_3}}{\a_1 \a_2} F^{3}_{12} \right) - \frac {g}{6} e^{-\Phi}(2e^{2\l_1} - e^{2\l_2} - e^{2\l_3}) \, ,
\label{lambdaa}
\ee
and cyclic in the $1,2,3$ indices for the other two equations.
Also we have denoted the field strength components by $F^i_{jk}$ 
in the $\s^j\wedge \s^k$ basis. 
In terms of the gauge field components they read 
\be 
F^{1}_{23}=A^1_1+g A^2_2 A^3_3\ , \qq {\rm and\ cyclic\ permutations}\ .
\label{fijk}
\ee
From the gravitino equation one determines the warp factor $f$ in terms of the
dilaton $\Phi$ as 
\be 
f = {\Phi\ov 3}\ , 
\label{fphi}\ 
\ee
as well as the differential equation 
\be 
{1\ov \a_1} {d\a_1\ov d\rho} 
=  \frac {e^{\Phi}}{6} \left( e^{\l_1} \frac {F^1_{23}}{\a_2 \a_3} - 
5 e^{\l_2}\frac {F^2_{31}} {\a_3\a_1} - 
5 e^{\l_3}\frac {F_{12}^3}{\a_1\a_2} \right) 
+ \frac {g}{24} e^{-\Phi} (e^{2\l_1} + e^{ 2 \l_2} + e^{2 \l_3}) \ ,
\label{ai}
\ee
together with two more equations obtained by cyclic permutations of the $1,2,3$ indices. 
Furthermore, from $\delta \psi_{\rho}$ we can obtain the radial dependence 
of the spinor $\epsilon$, which is simply given by
\be
\e = e^{f/2} \e_0 = e^{\Phi/6} \e_0 \, ,
\label{spinor}
\ee 
for $\e_0$ a constant spinor obeying the projection conditions 
(\ref{projections}). This radial dependence is 
of the general form $\e = g_{00}^{1/4} \e_0$, 
which can be proved using general arguments based on the 
supersymmetric algebra. The dependence on the particular model is only 
via the projections imposed on the constant spinor $\e_0$, which 
reduce the number of its independent components (see for instance \cite{Kallosh}).
  
Using the appropriate formulae in  \cite{Salam} we may lift our 8-dimensional
background into a full solution of 11-dimensional supergravity with only 
the metric turned on. The result is of the form $ds_{11}^2= ds^2_{1,3}+
ds^2_7$ where the 7-dimensional part is
\be
ds_7^2 = e^{-2\Phi/3} d\r^2 + e^{-2\Phi/3} \sum_{i=1}^3 \a_i^2 \s_i^2 +
e^{4\Phi/3} \sum_{i=1}^3 e^{2\l_i}(2/g \S_i + 2 A^i_i \s_i)^2\ .
\ee
This metric, when the various functions are subject to the conditions 
\eqn{A1} and \eqn{dilaton}-\eqn{ai}, describes $G_2$ holonomy manifolds
with an $SU(2)\times SU(2)$ isometry.

It is worth examining what the Killing spinor in \eqn{spinor} represents
from an eleven dimensional point of view.
Recall that, in general, when a supersymmetry variation parameter $\e$ 
is lifted
from eight to eleven dimensions, it is multiplied by a factor, 
i.e. $\e_{11}=e^{-\Phi/6} \e$.\footnote{This corrects an apparent typo in 
equation (34) of \cite{Salam}.}
Using, in our case, the expression \eqn{spinor} we see that the constant 
spinor 
$\e_0$ is indeed the 11-dimensional spinor which, being
subject to the projections \eqn{projections}, has 4 independent components.
We will next show that it splits into the form $\e_0=\e_{1,3} \times \e_7$
in such a way that the spinor $\e_7$ in  seven dimensions has only one independent 
component, in agreement with the correct amount of independent 
supercharges preserved by a
$G_2$ holonomy manifold. In order to proceed we specialize the index $\m$
to represent only the flat directions, i.e. $\m=0 ,1,2,3$
and we denote by $\bar\m=4,5,6,7$ the rest.
Then we may represent the gamma matrices in 11-dimensions as
\ba
&& \G^\m=\g^\m\times {\bf 1_4}\times {\bf 1_2}\ ,
\nonumber\\
&& \G^{\bar \m}=\g_5 \times \g^{\bar \m} \times {\bf 1_2}\ ,
\label{gg7}\\
&& \hat \G^i = \g_5 \times \bar \g_5 \times \tau_i\ ,
\nonumber
\ea
where $\g_5 =i \g^0\g^1\g^2\g^3$, $\bar\g_5 = \g^4\g^5\g^6\g^7$ and 
where we have used that $\e^{0123}=\e^{4567}=1$.
Using the split $\e_0=\e_{1,3} \times \e_7$ we see that the projections
\eqn{projections} imply 
\be
(\g_{ij}\times {\bf 1_2}) \e_7 = -({\bf 1_4}\times \tau_{ij})\e_7\ ,
\qq
(\g^7\times {\bf 1_2}) \e_7 = -i (\bar\g_5\times {\bf 1_2})\e_7\ .
\label{advv}
\ee
These are 8 conditions in total on the 8-component spinor $\e_7$ and 
therefore the latter has indeed only one independent component, as 
advertised. Moreover, as shown in footnote 6 below the spinor $\e_7$ is 
$G_2$ invariant. The spinor $\e_{1,3}$ is subject to no conditions
at all and therefore the $\cN=1$ supersymmetry in four dimensions is intact
we may have reduced supersymmetry if the Minkowski space is replaced by a 
Ricci flat manifold which admits less Killing spinors that Minkowski space).

For completeness we also construct the 3-form which is closed and 
co-closed and whose existence implies that the manifold has $G_2$ holonomy. 
On general grounds its components in the 7-bein basis $e^a\wedge e^b\wedge e^c$
are of the form  $\Phi^{(3)}_{abc}=i \bar \e_7 \G_{abc}\e_7$, 
where $a,b,c=1,2,\dots ,7$
and the gamma matrices in seven dimensions 
are the corresponding part of the decomposition 
\eqn{gg7}. Using the split $a=(i,\hat i,7)$, where $i=1,2,3$ and 
$\hi=i+3$, as well as the normalization choice
$i\bar \e_7 \G_{123}\e_7=1$, we find that 
\be
\Phi^{(3)}= {1\ov 6} \psi_{abc} \, e^a \wedge e^b \wedge e^c\ ,
\label{phi3}
\ee
where $\psi_{abc}$ are the octonionic structure constants with non-vanishing 
components in our basis being given by\footnote{In the same basis the
non-vanishing components of the 
$G_2$ invariant 4-index tensor $\psi_{abcd}$ are
\ba
\psi_{7ij\hk} = \e_{ijk}\ ,\qq \psi_{7\hi\hj\hk} =-\e_{ijk}\ ,\qq
\psi_{ij\hat m \hat n}=\d_{im}\d_{jn}-\d_{in}\d_{jm}\ .
\nonumber
\ea
Using these, one can show that the projectors \eqn{advv} imply that
\ba
(\G_{ab}+{1\ov 4} \psi_{abcd}\G_{cd})\e_7=0\ ,
\nonumber
\ea
which is precisely the condition for a $G_2$ invariant Killing spinor.}
\be
\psi_{ijk}=\e_{ijk}\ ,\qq \psi_{i\hj\hk}=-\e_{ijk}\ ,
\qq \psi_{7i\hj}=\d_{ij}\ .
\ee

It is convenient to cast the metric and the equations in the different form
\be
ds_7^2 = dr^2 + \sum_{i=1}^3 a_i^2 \s_i^2 +
\sum_{i=1}^3 b_i^2(\S_i + c_i \s_i)^2\ ,
\label{s7}
\ee
where $c_i = 2 A_i^{i}$ and 
\be
a_i=e^{-\Phi/3} \a_i\ , \qq b_i = e^{2\Phi/3} e^{\l_i} \ ,\qq e^{2\Phi}=b_1
b_2 b_3\ , \qq dr= e^{-\Phi/3}  d\rho \ .
\label{aibi}
\ee
Then the equations (\ref{dilaton}), (\ref{lambdaa}) and (\ref{ai}) become
\ba
{da_1\ov dr}& =& - \frac {b_2}{a_3} F^2_{31} - \frac {b_3}{a_2} F^3_{12}\ , 
\nonumber\\
{db_1\ov dr}& =& \frac {b_1^2}{a_2a_3} F^1_{23} -{g \ov 4 b_2 b_3 } 
(b_1^2 - b_2^2 - b_3^2) \, ,
\label{ba}
\ea
and cyclic in the $1,2,3$ indices, where 
the field strength components in \eqn{fijk} are computed using 
\ba
A_1^1 & = & \frac {a_1}{g} 
\Big[ - \frac {a_2^2+a_3^2-a_1^2}{2a_1a_2a_3} \frac {b_2^2+b_3^2}{2b_2b_3} +
\frac {b_3^2-b_1^2}{2b_3b_1} 
\frac {b_2}{b_1} \frac {a_3^2 + a_1^2 - a_2^2}{2a_1a_2a_3} - 
\frac {b_1^2-b_2^2}{2b_1b_2} \frac {b_3}{b_1} 
\frac {a_1^2 + a_2^2 - a_3^2}{2a_1a_2a_3} \Big] \nonumber \\ 
\,    & = & -{1\over g} \frac {d_2^2 + d_3^2 -d_1^2}{2d_2d_3}
\equiv - \frac {1}{g} \Omega^{23}_1\ ,
\label{A11}
\ea
where $d_i \equiv \frac {a_i}{b_i}$ and cyclic in $1,2,3$.
We see that the generalized twist condition \eqn{A1} takes the form 
of the ordinary twist, but for an auxiliary 3-sphere deformed metric 
obtained by replacing the $a_i$'s in the metric \eqn{s7} 
by the $d_i$'s defined above.
In the rest of this paper we will set the parameter $g=2$ which is equivalent 
to the rescaling $b_i\to b_i g/2$. This does not apply for the various formulae
in the appendix.

It is worth examining the limit where the radius of the ``spacetime'' 
3-sphere becomes very large so that it can be approximated by $\IR^3$. 
This means that effectively the D6-branes are unwrapped. This limit 
can be taken systematically as follows: consider the rescaling $\s_i\to \e 
dx_i$, $b_i\to \e b_i$ and $r\to \e r$ in the limit $\e\to 0$. 
Then, since the functions $c_i=2 A^i_i$ do not scale, the metric 
\eqn{s7} takes the form $ds^2_7=dx_i^2 + ds^2_4$, where the four-dimensional
non-trivial part of the metric is
\be
ds_4^2 = dr^2 + \sum_{i=1}^3 b_i^2 \S_i^2\ .
\label{EEHH}
\ee 
The coefficients $b_i$ as functions of $r$ obey a set of differential 
equations that also follow from the above mentioned limiting procedure
from \eqn{ba}. Indeed the first equation in \eqn{ba} reduces to the statement
that the coefficients $a_i=\hbox {constant}$ and therefore they can be absorbed into 
a rescaling of the new coordinates $x_i$, as we have already done above.
The result is 
\be
{db_1\ov dr}= {1\ov 2 b_2 b_3}(b_2^2+b_3^2-b_1^2)\ ,
\qq {\rm and\ cyclic\ permutations}\ .
\label{grt}
\ee
This is nothing but the Lagrange system or, equivalently, 
the Euclidean version of the Euler spinning top system.
The four-dimensional metrics \eqn{EEHH} governed by that system 
correspond to a class of hyperk\"ahler 
metrics with $SU(2)$ isometry with famous example, when an extra $U(1)$ 
symmetry develops (i.e., for instance when $b_2=b_3$), the 
Eguchi--Hanson metric which is the first non-trivial ALE 
four-manifold.\footnote{In fact, the Eguchi--Hanson metric is the 
only regular metric in the family described by \eqn{grt}.
As it was shown in \cite{belinskii} a generalization of it with $b_1\neq
b_2\neq b_3\neq b_1$ leads to singular metrics. It can be shown that, from a 
string theoretical view point, this corresponds
to continuous distributions of D6-branes in type IIA with physically 
unacceptable densities.}
This is in agreement with the fact that the near horizon limit of D6-branes of 
type IIA when uplifted to M-theory, contains, besides the D6-brane 
worldvolume, the Eguchi--Hanson metric.

Returning back to the generic case, it is obvious that integrating 
the system of first order non-linear equations \eqn{ba}
is a difficult
task in general. Nevertheless one can show that 
\be
I= a_1a_2a_3 - a_1 b_2 b_3 c_2 c_3 - a_2 b_3 b_1 c_3 c_1 - a_3 b_1 b_2 c_1 c_2
\ ,
\label{intt}
\ee
is a constant of motion. The existence of this constant of motion fits well
with the fact that the 3-form in \eqn{phi3}, after using the explicit 
basis \eqn{ba31} below in terms of the $SU(2)$ Maurer--Cartan 1-forms,
can be written as 
\be
\Phi^{(3)}= I \s_1\wedge \s_2 \wedge \s_3 + d\L\ ,
\ee
where $I$ is the conserved quantity in \eqn{intt} and $\L$ is some 2-form.
Hence the conservation of $I$ is a direct consequence of the closure of the
3-form $\Phi^{(3)}$, and appears as the coefficient of the volume form of the 
``spacetime'' 3-sphere. Notice that there is no conserved quantity 
associated with the internal 3-sphere.\footnote{In the notation 
of \cite{Brandhuber} $p=I $ and $q=0$. In principle, the information contained
into our system \eqn{ba} for the metric \eqn{s7} is also encoded into 
equations (80)-(81) of \cite{Brandhuber} for the metric (78)-(79) of the same 
reference. These equations are highly non-linear second order equations for three functions. In our 
approach they would arise upon eliminating three
among our six unknown functions. A simple counting argument shows that in both
cases the number of integration constants is the same.
We note here that it does not seem possible to investigate metrics with
both $p\neq 0$ and $q\neq 0$ using eight-dimensional gauged supergravity.
The reason is that, in the original metric ansatz \eqn{s8} there cannot be 
by definition any dependence on the internal $SU(2)$ coordinates that 
parameterize the Maurer--Cartan 1-forms $\S_i$.}
A promising avenue towards finding
explicit new solutions will arise if the system \eqn{ba} can be related to 
well studied in the literature spinning top-like systems which in many cases
are integrable.
This is the line of approach advocated in \cite{Sfetsos}, 
but will leave this and 
related investigations for future research.

Let us now consider the consistent truncation $a_2=a_3$ and $b_2=b_3$,
where an extra $U(1)$ symmetry develops.
Then after some algebra we conclude that the remaining four independent 
functions obey the system\footnote{It is straightforward to 
verify that the further consistent truncation with $a_1=a_2=a_3$ 
and $b_1=b_2=b_3$ gives a system which is trivially solved, 
leading to the metric of \cite{BS}.}
\ba
\dot a_1& =& {1\ov 4} {a_1^3 b_2^4\ov a_2^4 b_1^3}\ ,
\nonumber\\
\dot a_2 & = & {1\ov 2} {b_1\ov a_2} -{3\ov 8} {a_1^2 b_2^2\ov
a_2^3b_1} + {1\ov 8} {a_1^2 b_2^4\ov a_2^3 b_1^3}\ ,
\nonumber\\
\dot b_1 & = & -{1\ov 2} {b_1^2\ov a_2^2} + {3\ov 8} {a_1^2 b_2^2\ov a_2^4}
-\ha \left({b_1^2\ov b_2^2}-2\right)\ ,
\\
\dot b_2 &  = & 
\ha {b_1\ov b_2} -{1\ov 8} {a_1^2 b_2^5\ov a_2^4 b_1^3} \ ,
\nonumber
\ea
where we have used that in this case $c_2=-{a_1 b_2\ov 2 a_2 b_1}$ and
$c_1=2 c_2^2-1$.
This system coincides (after we let $r\to -r$)
with that in equation (23) of \cite{Cvetic} and in the limit 
of $a_1=0$ it is just the system corresponding to the resolved conifold.

We will finally show how the system of equations (\ref{ba}) 
can also be derived from self-duality 
of the spin connection for the 
seven manifold.
In order to do so, we will split the 
indices in (\ref{s7}) as before, namely as $a=(i,\hi,7)$, and use the 
7-bein basis
\be e^7=dr\ , \qq e^i = a_i \s_i\ , \qq e^\hi = b_i(\S_i+c_i\s_i)\ ,\qq
i=1,2,3 \ , \quad \hi=i+3\ .
\label{ba31}
\ee
We then compute
\ba 
de^i & = & {\dot a_i\ov a_i} e^7 \wedge e^i \ 
+\ \ha {a_i\ov a_j a_k}\ \e_{ijk}\
e^j\wedge e^k \ ,
\nonumber\\
de^\hi & = & {\dot b_i\ov b_i} e^7 \wedge e^\hi \ +\ 
{ b_i \dot c_i\ov a_i}\ e^7 \wedge e^i 
\nonumber\\
&& +\ \ha b_i\ \e_{ijk} \left({1\ov b_j b_k}\ e^\hj\wedge e^\hk \ +\
{c_i+c_jc_k\ov 
a_j a_k}\ e^j\wedge e^k \ -\ 2 {c_j\ov a_j b_k}\  e^j\wedge e^\hk \right)\ ,
\ea
where the dot stands for $\frac {d \:}{dr}$. Using then the Cartan's structure 
equations $de^a+ \om^{ab}\wedge e^b=0$ we compute the spin connection
\ba
\om^{i7}& = &{\dot a_i\ov a_i}\ e^i \ + \ {b_i \dot c_i\ov 2 a_i}\ e^\hi\ ,
\nonumber\\
\om^{\hi 7}& = &{\dot b_i\ov b_i}\ e^\hi \ + \ {b_i \dot c_i\ov 2 a_i}\ e^i\ ,
\nonumber\\
\om^{ij} & =& \ha \e_{ijk}\left({a_i\ov a_j a_k} + {a_j\ov a_i a_k} -
{a_k\ov a_i a_j}\right) e^k 
\ -\ \ha \e_{ijk}\ {b_k\ov a_i a_j} (c_k+c_i c_j)\ e^\hk\ ,
\\
\om^{\hi\hj} & = & 
\ha \e_{ijk}\left({b_i\ov b_j b_k} + {b_j\ov b_i b_k} -
{b_k\ov b_i b_j}\right) e^\hk 
\ -\ \ha \e_{ijk} \frac {c_k}{a_k} \left({b_i \ov b_j} + {b_j \ov b_i}\right) 
\ e^k\ ,
\nonumber\\
\om^{i\hj}& = & {b_i \dot c_i\ov 2 a_i}\d_{ij}\ e^7 \ + \ \ha\e_{ijk}\ 
{b_j\ov a_i a_k}\ (c_j+c_k c_i)\ e^k \ - \  \ha \e_{ijk} \frac {c_i}{a_i} 
\left({b_j\ov b_k}-{b_k\ov b_j}\right)\ e^\hk\ .
\nonumber
\ea
  
Then, let us recall that imposing the self-duality condition on the spin 
connection, i.e. $\om^{ab}=\ha \psi^{abcd} \om^{cd}$, where $\psi^{abcd}$
is the $G_2$ invariant 4-index tensor, is equivalent in our basis
to the following seven equations 
\ba
\om^{7i} & = & \e_{ijk} \om^{j \hk} \, , \nonumber \\
\om^{7\hi} & = & \frac {1}{2} \e_{ijk} (\om^{jk}-\om^{\hj \hk}) \, , 
\label{spth}\\
\om^{i \hi} & = & 0 \ . 
\nonumber 
\ea
Applying these to our case we obtain the differential equations (\ref{ba}) and
the generalized twist condition \eqn{A11}, 
plus the condition $\sum_{i=1}^3 \frac {b_i}{a_i} \dot{c}_i =0$, which 
is equivalent to (\ref{constraint}) in the appendix and is satisfied 
automatically once \eqn{ba} and \eqn{A11} are.
Since self-duality of the spin connection in seven dimensions 
implies that the 3-form defined in 
\eqn{phi3} is closed and co-closed and, therefore, $G_2$ holonomy 
(noted in \cite{AL,Floratos}, proved explicitly in \cite{Bilal} and used to
rederive the metric of \cite{BS} in \cite{BP})
we have shown that our equations \eqn{ba} (or equivalently (\ref{dilaton}), 
(\ref{lambdaa}) and (\ref{ai})) indeed describe a manifold of $G_2$ holonomy.

It will be interesting to extend the eight-dimensional gauged supergravity 
approach to $G_2$ manifolds in order to 
find general conditions for manifolds with weak $G_2$ holonomy \cite{Gray}
having 
an $SU(2)\times SU(2)$ isometry. The main difference in this case is that 
the three form is no longer closed, i.e. it obeys 
$d\Phi^{(3)}\sim *\Phi^{(3)}$ and consequently the Minkowski
metric $ds^2_{1,3}$ has to be replaced by an Einstein space with negative
cosmological constant. Nevertheless, supersymmetry can be preserved 
and a generalization of the self-duality condition on the spin connection 
\eqn{spth} leading to manifolds with weak $G_2$ holonomy also exists
\cite{Bilal}.
We also believe that the eight-dimensional approach to $G_2$ manifolds 
will also prove useful in the investigation of Spin(7) manifolds.
We hope to report work along these lines in the future.


\bs\bs

\centerline {\bf Acknowledgments}

R. H. acknowledges the financial support provided through the European
Community's Human Potential Programme under contract HPRN-CT-2000-00131
Quantum Spacetime and by the Swiss Office for Education and Science and the Swiss National Science Foundation, 
and hospitality of the Erwin Schr\"odinger Institute in Vienna. 
K. S. acknowledges the hospitality and the financial support of the Institute 
of Physics at the University of Neuch\^atel 
in which he was a member while part of this work was done.


\appendix

\renewcommand{\theequation}{\thesection.\arabic{equation}}
\csname @addtoreset\endcsname{equation}{section}
  
\section{Appendix}

In this appendix we will provide some details on the derivation of the Killing 
spinor equations and conditions \eqn{twist}-\eqn{spinor}.
  
As already noted, the only consistent way to obtain a consistent 
set of differential equations
from the 
supersymmetry variations (\ref{susy}) and \eqn{susyg} is to impose projections 
(\ref{projections}) on the spinors. 
We also provide for convenience the expressions for $P_{\m \, ij}$ and $Q_{\m \, ij}$
defined in \eqn{pq}. For the diagonal matrix $L^i_\a$ in \eqn{lia}, it is convenient to 
represent them as forms in the index $\mu$,
\be
P_{ij}=\pmatrix{\del \l_1 & g A^3 \sinh \l_{12} & g A^2 \sinh\l_{31}\cr
g A^3 \sinh\l_{12} & \del \l_2 & g A^1 \sinh\l_{23} \cr
g A^2 \sinh \l_{31} & g A^1 \sinh\l_{23} & \del \l_3 }\
\ee
and 
\be
Q_{ij}=\pmatrix{0 & -g A^3 \cosh \l_{12} & g A^2 \cosh\l_{31}\cr
g A^3 \cosh\l_{12} &0 & -g A^1 \cosh\l_{23} \cr
-g A^2 \cosh \l_{31} & g A^1 \cosh\l_{23} & 0 }\ .
\ee

We start with the $i=1$ case in the gaugino equation, 
$\delta \chi_1=0$, which implies two different equations: factorizing 
$\hat{\G}_2 \hat{\G}_3 \, \e$ we get
\be 
\frac {d\l_1}{d \rho} + \frac {2}{3} \frac {d\Phi}{d \rho} = 
e^{\Phi + \l_1} \frac {F^1_{23}}{\a_2 \a_3} - 
\frac {g}{4} e^{-\Phi} ( e^{2\l_1} - e^{2\l_2} -e^{2\l_3}) \ ,
\label{a.1}
\ee
where $F^1_{23}$ is defined in \eqn{fijk}. 
In addition, from terms proportional 
to $\hat{\G}_2 \G_3 \, \e$ we get 
\be
e^{\Phi+\l_1} \frac {F^1_{\r1}}{\a_1} + g \left( \frac {A^3_3}{\a_3} 
\sinh \l_{12} - \frac {A^2_2}{\a_2} \sinh \l_{31} \right)=0 \ ,
\label{a2}
\ee
where $F^1_{\r1} = \del_\r A^1_1$,
or equivalently, after the change of variables \eqn{aibi} (and setting $g=2$)
\be
{b_1 \dot c_1\ov a_1}+ {c_3\ov a_3}\left({b_1\ov b_2} - {b_2\ov b_1}\right)
- {c_2\ov a_2}\left({b_3\ov b_1} - {b_1\ov b_3}\right)= 0\ .
\ee
The four additional equations corresponding to $\d \chi_i=0$,
for $i=2,3$ can be 
obtained from \eqn{a.1} and \eqn{a2} by cyclic permutations in the indices 
$1,2,3$.
Using then the constraint $\l_1 + \l_2 + \l_3 =0$ we get
\be
{d\l_1\ov d\rho} = \frac {e^{\Phi}}{3} 
\left( 2 \frac {e^{\lambda_1}}{\a_2 \a_3} F^{1}_{23} - 
\frac {e^{\lambda_2}}{\a_3\a_1} F^2_{31} - 
\frac {e^{\lambda_3}}{\a_1 \a_2} F^{3}_{12} \right) 
- \frac {g}{6} e^{-\Phi}(2e^{2\l_1} - e^{2\l_2} - e^{2\l_3}) \
\label{ej1}
\ee
and
\be
{d\Phi\ov d\rho} = {1\ov 2} e^\Phi \left( \frac {e^{\lambda_1}}{\a_2 \a_3} F^{1}_{23} + \frac {e^{\lambda_2}}{\a_3\a_1} F^2_{31} 
+ \frac {e^{\lambda_3}}{\a_1 \a_2} F^{3}_{12} \right)
+ {g\ov 8} e^{-\Phi} 
(e^{2\l_1} + e^{2\l_2} + e^{2\l_3})\ .
\label{ej2}
\ee
  
Next we turn to the gravitino variation, $\delta \psi_{\m}=0$, where 
we should distinguish 
two possibilities, according to whether $\mu$ is a 
coordinate on the wrapped 3-sphere or on the unwrapped directions. 
Thus, if $\mu=\s_1$, from $\hat{\G}_2 \hat{\G}_3 \, \e$ we get
\be
\om_1^{23} + g A_1^1 \cosh \l_{23} - \a_1 
\frac {e^{\Phi}}{6} \left( \frac {e^{\l_2}}{\a_2} F^2_{\r2} 
+ \frac {e^{\l_3}}{\a_3} F^3_{\r3} - 
5 \frac {e^{\l_1}}{\a_1} F^1_{\r1} \right)=0
\label{a4}
\ee
and, from terms proportional to $\hat{\G}_2 \G_3 \, \e$
\be
\frac {1}{\a_1} \frac {d \a_1}{d \rho} 
- \frac {1}{6} e^{\Phi} \left( \frac {e^{\l_1}}{\a_2 \a_3} F^1_{23} -  
5 \frac {e^{\l_2}}{\a_3 \a_1}{F^2_{31}} 
- 5 \frac {e^{\l_3}}{\a_1 \a_2} F^3_{12} \right) - 
\frac {g}{24} e^{-\Phi} (e^{2\l_1}+e^{2\l_2}+e^{2\l_3}) =0 \, ,
\label{a5}
\ee
where we have used the spin connection component 
$\om^{i \rho}_i = \frac {d \a_i}{d \rho}$ for the metric 
ansatz (\ref{s8}). The equations obtained from cyclicity 
of (\ref{a4}) and (\ref{a5}) correspond to the choices $\mu = \s_2, \, \s_3$. 
The generalized twist (\ref{twist}) 
is then derived from (\ref{a2}) and (\ref{a4}). It amounts to turning on a 
gauge field given by (\ref{A1}). 

It is important to verify that substituting
back \eqn{A1} into \eqn{a2} and after using \eqn{ej1}, \eqn{ej2} and
\eqn{a5},
gives no new constraints for the various functions. After a straightforward
but lengthly computation, one can show that this is indeed the case.

If $\mu$ is a coordinate on the unwrapped part of the worldvolume, using 
$\om^{\mu \rho}_{\mu} = e^f \frac {df}{d \rho}$ and comparing 
terms proportional to $\G_{\mu} \G_7 \, \e$ 
we get and equation relating the warp factor and the dilaton as
\be 
\frac {d f}{d \rho} = \frac {1}{3} \frac {d \Phi}{d \rho}\ .
\ee
This then leads to \eqn{fphi} in the main text (a possible constant of 
integration can be absorbed into a rescaling of the corresponding unwrapped
coordinates). Also from comparison of terms proportional 
to $\G_\m \G_7 \G_1\hat \G_1\e$ we obtain a constraint for the field strength 
\be
e^{\l_1} \frac {F^1_{\r1}}{\a_1} + e^{\l_2} \frac {F^2_{\r2}} {\a_2} 
+ e^{\l_3} 
\frac {F^3_{\r3}} {\a_3} = 0 \ ,
\label{constraint}
\ee 
which holds identically from (\ref{a2}).
Finally, considering the gravitino equation for $x^\m=\r$ gives after some 
algebra a simple differential equation for the spinor $\e$ 
with solution given by \eqn{spinor}.
           


\end{document}